\documentclass[aps,prl,twocolumn,amsmath,amssymb,showpacs]{revtex4}
\usepackage{graphicx,color,dcolumn}
\usepackage{amssymb}
\usepackage{amsmath}
\usepackage{epsfig,color}
\usepackage{epstopdf}
\usepackage{xcolor}

\begin{document}

\title{Instabilities and relaxation to equilibrium in long-range oscillator chains}
\author {George Miloshevich${}^{1,2}$, Jean-Pierre Nguenang${}^{3,{4}}$,
Thierry Dauxois${}^{4}$, Ramaz Khomeriki${}^{1,5}$, Stefano
Ruffo${}^{5}$}

\affiliation{${\ }^{(1)}$Department of Physics, Faculty of Exact
and Natural Sciences, Tbilisi State University, 0128 Tbilisi,
Georgia\\
${\ }^{(2)}$Department of Physics, The University of Texas at
Austin, Austin TX 78712, USA \\
${\ }^{(3)}$ Fundamental Physics Laboratory: Group of Nonlinear
Physics and Complex Systems,
Department of Physics, University of Douala, P.O. Box 24157, Douala, Cameroon \\
${\ }^{(4)}$Laboratoire de Physique de l'ENS Lyon
, Universit\'e de Lyon, CNRS, 46, all\'ee d'Italie, 69007 Lyon, France \\
${\ }^{(5)}$ Dipartimento di Fisica e Astronomia and CSDC,
Universit\`a di Firenze, CNISM and INFN, via G. Sansone, 1, Sesto
Fiorentino, Italy}

\begin{abstract}
We study instabilities and relaxation to equilibrium in a long-range extension of
the Fermi-Pasta-Ulam-Tsingou (FPU) oscillator chain by exciting initially the lowest Fourier mode.
Localization in mode space is stronger for the long-range FPU model. This allows us to uncover the
{\it sporadic} nature of instabilities, i.e., by varying initially the excitation amplitude of the lowest
mode, which is the control parameter, instabilities occur in narrow amplitude intervals. Only for
sufficiently large values of the amplitude, the system enters a permanently unstable regime.
These findings also clarify the long-standing problem of the relaxation to equilibrium in the short-range
FPU model. Because of the weaker localization in mode space of this latter model, the transfer of energy is
retarded and relaxation occurs on a much longer time-scale.
\end{abstract}

 \maketitle

The relaxation to equilibrium in nonlinearly coupled oscillators
is {  still an open problem}~\cite{FPU,galavotti}. Starting with the
pioneering work of Fermi-Pasta-Ulam-Tsingou (FPU)~\cite{tod}, these
studies led to important discoveries in both statistical mechanics
and nonlinear science. In most cases, the analysis was restricted to
one-dimensional ($d=1$) lattices where oscillators interact only
with nearest neighbors, i.e. to short-range interactions. However, in
recent years there has been a growing interest in systems with
long-range interactions~\cite{long,Campabook}. In such systems,
either the two-body potential or the coupling at separation $r$
decays with a power-law $r^{-\alpha}$. When the power $\alpha$ is
less than the dimension of the embedding space $d$, these systems
violate additivity, a basic feature of thermodynamics, leading to
unusual properties like ensemble inequivalence, broken ergodicity,
quasistationary states.

The extension of the FPU problem to include long-range couplings is
rarely considered~\cite{flach,frac,helena}. Moreover, attention has
been focused mainly on finding conditions for the existence of
localized solutions like solitons or breathers. No one, to the best
of our knowledge, has tackled, in the context of long-range systems,
the original question posed by FPU on the time-scales for
relaxation to equilibrium when the energy is fed into the lowest Fourier mode. This is
the subject of this Letter. Moreover, we here show that the lessons learnt from the
long-range FPU model can be used to clarify key features of the original
short-range model.

Long-range coupled oscillator models have been previously introduced to cope
with dipolar interaction in mechanistic DNA models~\cite{dna}. They describe
also ferroelectric~\cite{electr} and magnetic~\cite{magnet}
systems, where the long-range coupling is provided again by dipolar forces.
Other candidates for application are cold gases: dipolar bosons~\cite{trom,atomdipol},
Rydberg atoms~\cite{rydberg}, atomic ions~\cite{kastner,ions}. Moreover, one can
mention optical wave turbulence~\cite{optics} and scale-free avalanche dynamics
\cite{ava}, where such long-range couplings appear.

In this Letter, we consider a generalization of the FPU model by
introducing a long-range coupling in the linear term, while keeping
the nonlinear term short-range. Choosing the power $1<\alpha\leq 3$,
the results do not depend much on the specific value. Dipolar systems correspond, as mentioned, to
the power $\alpha=3$, while the power $\alpha=2$ has been considered
for crack front propagation along disordered weak planes between
solid blocks \cite{ava} and contact lines of liquid spreading on
solid surfaces~\cite{liquid}.

Here, we repeat the FPU experiment by putting the energy initially
in the lowest Fourier mode. As for the short-range FPU model, an
exponential spectrum involving only odd modes forms on a short-time
scale~\cite{ruffo,flach1}. However, energy localization in Fourier
space is much stronger, as we will comment in the following. By increasing
the initial amplitude, a parametric instability sets in where the energy is
transferred to even modes~\cite{dri,poggi,chechin,christo}.

\begin{figure}[t]
\epsfig{file=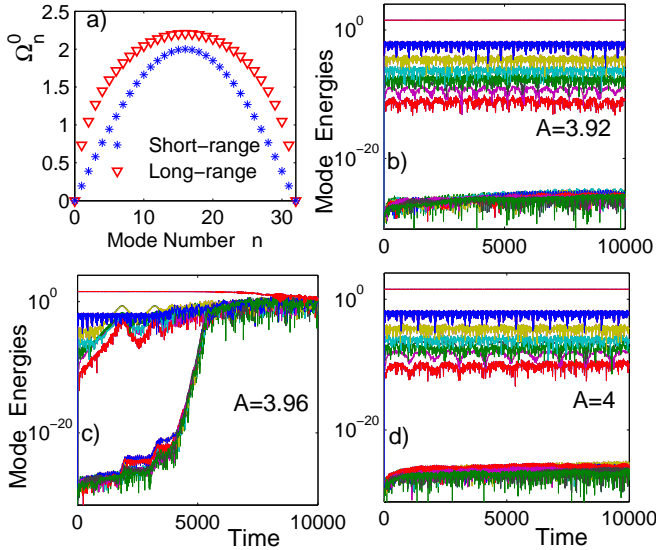,width=1\linewidth} \caption{a) Unperturbed
frequency distribution, $\Omega_n^0$ in formula \ref{disp0}, for the
short-range FPU $\alpha=\infty$ (stars) and for the long-range one
with $\alpha=2$ (triangles). b)-d) Time evolution of Fourier mode
energies obtained by numerically integrating Eqs.~\eqref{first}.
Initially, only the lowest mode is excited with increasing
amplitudes: $A=1.96$, $A=1.98$ and $A=2$ (graphs b), c) and d),
respectively). Warm colors (close to red) in b)-d) correspond to
even modes while cold colors (close to green) represent odd ones.
The number of oscillators in all panels is fixed to $N=32$ and the
long-range interaction power is $\alpha=2$.} \label{Fig1}
\end{figure}

For the long-range FPU this instability has a {\it sporadic} nature,
i.e. by increasing the amplitude one observes {\it instability
islands}, narrow amplitude intervals where instability sets in.
Similar {\it sporadic} instabilities (so called induction
phenomenon~\cite{ind}) have been observed in the short-range FPU
when exciting higher modes. As we will show below, these
instabilities play a much more important role in the long-range FPU
than in the short-range one. This is due to the strong localization
in mode space, which is ultimately determined by the non-equidistant
character of the unperturbed frequency spectrum, see
Fig.~\ref{Fig1}a. This instability drives the long-range system to
energy equipartition on a short time-scale. For the short-range FPU,
the simultaneous presence of {\it sporadicity} and weak localization
in mode space, which is due to the equidistant character of the
frequency spectrum shown in Fig.~\ref{Fig1}a, delays the convergence
to energy equipartition. Therefore, the study of these instabilities
for long-range systems clarifies an important aspect of the
relaxation to equilibrium in the traditional short-range FPU model.

Let us consider the following long-range Hamiltonian, which
describes a system of coupled nonlinear oscillators
\begin{equation}
\label{ham}
{\cal H} = \sum\limits_{{j=1}}^N \frac{\dot u_j^2}{2} +
\sum_{j>\ell=1}^N
{R_{j\ell}\frac{(u_j-u_\ell)^2}{2}}+\sum\limits_{j=1}^N
\frac{\left(u_{j+1}-u_j\right)^4}{4},
\end{equation}
with displacements $u_j$ and velocities $\dot u_j$ of the $N$
oscillators. Quartic nonlinearity has been chosen so that the
above model applies to the class of systems with inversion symmetry.
We define the algebraically decaying interaction matrix elements of the
harmonic term as
\begin{equation}
R_{j\ell} =  \frac{1}{{\lvert j-\ell\rvert}^\alpha} +
\frac{1}{(N-{\lvert j-\ell\rvert)}^\alpha},
\label{RJ}
\end{equation}
in which $1<\alpha\leq 3$ is a long-range interaction power. The
lower limit corresponds to a diverging harmonic interaction for
infinite $N$ while, beyond the upper limit we are in the short-range
regime~\cite{flach}. In order to be specific, we discuss here the case $\alpha=2$, but
the results of simulations are similar for all long-range powers~\cite{second}.

Hamilton's equations of model~(\ref{ham}) are
\begin{equation} \label{first}
\ddot{u}_\ell = \sum_{j=1}^N
{R_{j\ell}}(u_j-u_\ell)+\left(u_{\ell+1}-u_\ell\right)^3+\left(u_{\ell{
-}1}-u_\ell\right)^3.
\end{equation}
In momentum representation
\begin{equation} \label{moment}
{Q}_{n} = \frac{1}{N}\sum_{\ell=1}^Nu_\ell \,e^{iq_n \ell},
\end{equation}
where $n$ is the normal mode number, and $q_n=2\pi n/N$, one can rewrite
\eqref{first} in the following form
\begin{equation} \label{first1}
\ddot{Q}_{n} = -\left(\Omega_n^0\right)^2Q_n-\sum_{\langle
m,r,p\rangle}^N G_{mrp}^n Q_mQ_rQ_p,
\end{equation}
where normal mode numbers satisfy the condition $m+r+p-n=0,N,2N,...$
and $\langle m,r,p\rangle$ further restricts the sum so that each
triplet of modes $\{m,r,p\}$ is present only once.
$G_{mrp}^n=D_{mrp}\omega_m\omega_r\omega_p\omega_n$ are mode
coupling constants where $\omega_r\equiv 2\sin(q_r/2)$ and $D_{mrp}$
stand for the number of permutations of the set $\{m,r,p\}$, while
$\Omega_n^0$ are unperturbed frequencies of the long-range normal
modes
\begin{equation}
\Omega_n^0 = \sqrt{2\sum_{m=1}^N \frac{1-\cos(q_n m)}{m^\alpha}}.
\label{disp0}
\end{equation}
These frequencies are plotted in Fig.~\ref{Fig1} for both a long-range
and a short-range case. As mentioned above the spectrum is not equidistant
in the long-range case.
Originally, the FPU paradox involved exciting the first Fourier mode
$n=1$ and looking for the equipartition of energy with respect to
all other modes. Thus, we consider the following initial condition
\begin{equation}
u_\ell = A\left[e^{i(q_1\ell-\Omega_1^0 t)}+c.c.\right]
\label{plane0}
\end{equation}
which implies that energy is carried by normal modes
$Q_1=A\,e^{i\Omega_1^0t}$ and $Q_{N-1}=\left(Q_{1}\right)^*$. In
what follows we will study the evolution and instability of this
initial excitation.

First of all, let us note that this mode could be treated, in the
first approximation, as a source field and causes a nonlinear
frequency shift to all normal modes. Indeed, examining nonlinear
terms in~\eqref{first1} for $Q_n$ normal mode evolution containing
$G_{n,1,N-1}^nQ_nQ_1Q_{N-1}$, we conclude that the expression for
renormalized frequencies is
\begin{equation}
\Omega_{n} = \sqrt{\left(\Omega_{n}^0\right)^2+
G_{n,1,N-1}^n\left|Q_1\right|^2}. \label{disp11}
\end{equation}
In addition, this source mode excites the branch of odd modes (often
referred to as normal mode bush or q-breather~\cite{chechin,christo,flach1}).
For instance, the initially absent $Q_3$ mode is generated according to the perturbation
analysis of~\eqref{first1} $\ddot{Q}_{3} =
-\left(\Omega_3^0\right)^2Q_3-G_{111}^3 \left(Q_1\right)^3$ giving
the exact solution
\begin{equation}
Q_3=\frac{G_{111}^3A^3}{9\Omega_1^2-\Omega_3^2}
\left[e^{3i\Omega_1t}-\cos(\Omega_3t)-
i\frac{3\Omega_1}{\Omega_3}\sin(\Omega_3t)\right].
\label{pert}
\end{equation}
Thus, in addition to the triple source harmonic $3\Omega_1$, $Q_3$
acquires the harmonic pair $\pm\Omega_3$ as well. In the long-range
case, the perturbation analysis is well justified since low
frequencies $\Omega_j$ are not equidistant, i.e. the ratio
$(3\Omega_1 -\Omega_3)/\Omega_1$ remains finite even in the
$N\rightarrow\infty$ limit, in contrast to the short-range FPU
system. This itself guarantees the perturbative scaling
$\left|Q_1\right|\gg \left|Q_3\right|$ and, as we will see below,
leads to an excellent agreement between analytical predictions and
numerical results. Proceeding further, we can derive perturbatively
the expression for $Q_5$ with harmonics $\pm \Omega_5$,
$\pm\Omega_3+2\Omega_1$ and $5\Omega_1$, and then for all odd normal
modes which have exponentially decaying amplitudes with increasing
mode number. We define such a steady distribution set of
multicomponent and multi-frequency odd modes as $Q_r^{(a)}$, where
$(a)$ indicates different harmonics of the $r$-th odd mode. Further,
we should discuss how to derive the instability properties of this
group of lowest energy modes.


First of all, we note that the initially excited lowest mode
$Q_1$ alone (when all other modes are not excited) is parametrically stable in
both long and short-range cases: thus, in order to analyze the
instability process, it is necessary to consider the combination of
odd modes created by the initial excitation of the first mode.
Following a parametric instability approach, we consider pair of
equations from the set~\eqref{first1}
\begin{eqnarray} \label{first2}
\ddot{Q}_{n} &=& -\left(\Omega_n\right)^2Q_n-G_{mrp}^n
Q_mQ_r^{(a)}Q_p^{(b)} \\
\ddot{Q}_{m} &=& -\left(\Omega_m\right)^2Q_m-G_{nrp}^m
Q_n\left(Q_r^{(a)}\right)^*\left(Q_p^{(b)}\right)^*, \nonumber
\end{eqnarray}
where we have restricted the sum to mode harmonics such that the
relative frequency
\begin{equation}
\Delta=\Omega_r^{(a)}+\Omega_p^{(b)}-\Omega_m-\Omega_n,
\label{shift}
\end{equation}
\begin{figure}[t]
\epsfig{file=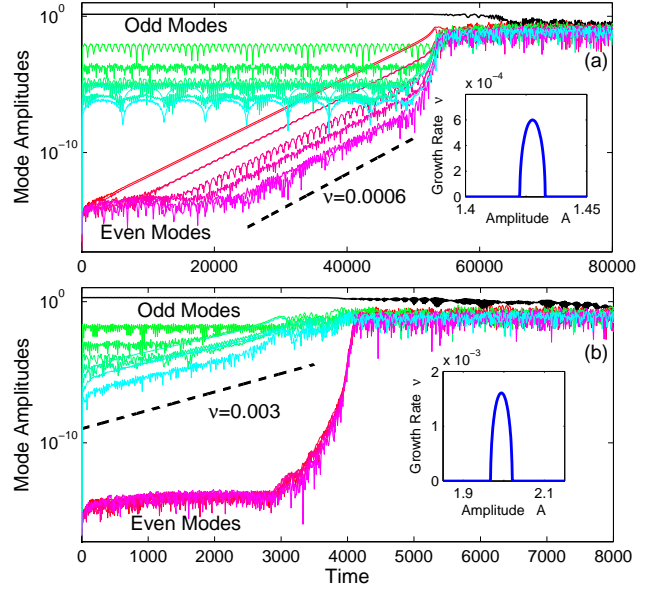,width=1\linewidth} \caption{ Long-range
dynamics: Time dynamics of normal modes from numerical simulations
on Eqs.~\eqref{first} with initial mode amplitude $A=1.4225$ (graph
a) and $A=1.97$ (graph b). {Curves are color-coded according to the
same principle as} in Fig.~1. Insets show the dependence of growth
rates on the initial lowest mode amplitude according to the
theoretical formula \eqref{grow}. Dashed lines show the estimated
grow rates for even (a) and odd (b) modes.} \label{Fig2}
\end{figure}
is close to zero. Then, introducing the transformations $Q_{n}=
{\cal Q}_{n}e^{i\Omega_{n}t}$ and $Q_{m}= {\cal
Q}_{m}e^{-i\Omega_{m}t}$, we get
\begin{eqnarray} \label{first3}
\ddot{\cal Q}_{n}+2i\Omega_n\dot{\cal Q}_n&=&-G_{mrp}^n
{\cal Q}_m{\cal Q}_r^{(a)}{\cal Q}_p^{(b)}e^{i\Delta t} \\
\ddot{\cal Q}_{m}-2i\Omega_m\dot{\cal Q}_m &=& -G_{nrp}^m {\cal
Q}_n\left({\cal Q}_r^{(a)}\right)^*\left({\cal
Q}_p^{(b)}\right)^*e^{-i\Delta t} \nonumber.
\end{eqnarray}
The instability takes place when the approximate resonance
condition $\Delta \approx 0$ holds. We can compute more precisely
this threshold assuming a slow time evolution in~\eqref{first3}
and neglecting second order time derivatives there. Thus, seeking for
solution of the form
\begin{equation}
{\cal Q}_n=F_1e^{(\nu+i\Delta/2)t} \quad \mbox{{and}}\quad {\cal
Q}_m=F_2e^{(\nu-i\Delta/2)t}, \label{solu}
\end{equation}
with $F_1$, $F_2$ constants and real growth increment $\nu$, we get
two coupled algebraic equations
\begin{eqnarray}
\Omega_n(\Delta{-}2i\nu)F_1 &-&G_{mrp}^n
{\cal Q}_r^{(a)}{\cal Q}_p^{(b)}F_2=0 \label{ageb}\\
\Omega_m(\Delta{+}2i\nu)F_2 &-&G_{nrp}^m \left({\cal
Q}_r^{(a)}\right)^*\left({\cal Q}_p^{(b)}\right)^*F_1=0.  \nonumber
\end{eqnarray}
The solvability condition of this system leads to the growth rate
\begin{equation}
\nu=\frac{1}{2}\sqrt{G_{mrp}^n G_{nrp}^m \left|{\cal
Q}_r^{(a)}\right|^2\left|{\cal
Q}_p^{(b)}\right|^2\bigl/{(}\Omega_n\Omega_m{)}-\Delta^2}.
\label{grow}
\end{equation}
Usually the first term under the square root is very small but the
growth rate could still be real (inducing the instability) if the
resonance condition $\Delta\rightarrow 0$ is realized. Thus,
changing $A$, we monitor the variation of $\Delta$ given
by~\eqref{shift}. We check for which set of the renormalized
frequencies of the modes the resonance condition is fulfilled. As
long as $r$ and $p$ are both odd, the resonance condition can be
realized separately for odd and even modes. In other words, for some
value of $A$, only even modes become unstable and their amplitude
grow, while odd modes remain at their stationary values (which they
abruptly acquire after the initial interaction with $Q_1$ mode).
Only when even modes grow sufficiently, the odd modes instability
develops. However, this is not always the case: If only odd modes
participate in the resonance condition, they will first leave their
stationary vlaues and grow exponentially, while even modes will join
only after odd modes reach the nonperturbative limit.

It is very important to stress that the instabilities we observe are
{\it sporadic}, which means that if the instability is realized for
certain values of $A$, it will not remain for larger values of $A$,
since the resonance condition will be violated. This is clearly seen
in numerical simulations [see Fig. \ref{Fig1}b)-d)] where the
dynamics of mode energies computed from the relation
$E_j=N\left(\left|\dot
Q_j\right|^2+\Omega_j^{2}\left|Q_j\right|^2\right)\Bigl/2$ is shown.
As shown in Fig.~\ref{Fig1}b)-d), the instability  appears for
$A=1.98$, allowing the onset of equipartition in the system where
all (time averaged) mode energies become equal. However, a further
increase of $A$ leads again to the stable q-breather state. Only for
sufficiently large amplitudes ($A>2.6$) the system is always
unstable.

Let us note that only the instability of one pair of even (odd)
modes is sufficient for the development of exponential growth of
other even (odd) modes with the same growth rate. Indeed, even (odd)
modes are coupled via the following mechanism
\begin{equation}
\label{first4}
\ddot{Q}_{n+2} = -\left(\Omega_{n+2}\right)^2Q_{n+2} -G_{n11}^{n+2}
Q_nQ_1Q_1,
\end{equation}
derived from Eq.~\eqref{first1}.

From the above analytic considerations, it is evident that in order to
find the instability islands, one should analyze the
effective frequency difference~\eqref{shift}. It is clear that when
it vanishes, the growth rate~\eqref{grow} becomes real and the
instability takes place. In other words, one has to find the set of
modes $m,n,r,p$ and the amplitude $A$ such that $\Delta\rightarrow 0$.
For our initial condition, it is natural to seek resonances with the primary harmonic
$\Omega_1$ and perhaps the harmonics of the third mode $\Omega_3^{(a)}$, which can be
more easily excited.

We have found multiple sets of these parameters for which either odd
or even mode instabilities occur. For instance, if one chooses the
set  $n=6$, $m=2$, $r=1$, $p=3$ with frequencies $\Omega_6$,
$\Omega_2$, $\Omega_1$, $3\Omega_1$, respectively, one gets
$\Delta=0$ for $A=1.425$. For slightly smaller or larger values of
$A$, the resonance condition is violated and the system becomes
stable again. It is possible to check these theoretical predictions
via direct numerical simulations of~Eqs.~\eqref{first} and initial
condition~\eqref{plane0} with $A=1.4225$. The results are displayed
in Fig.~\ref{Fig2}(a), where the inset shows the dependence of
growth rate calculated from~\eqref{grow} on $A$, while numerical
simulations are presented in the main plot. It is seen that even
modes grow first with growth rate $\nu=0.0006$ which is in good
agreement with the theoretical value. It is remarkable that the first
modes to participate in the growth almost in unison are indeed
$\Omega_6$ and $\Omega_2$.

Furthermore, it is possible to find a scenario where odd
modes grow first. This is realized for the set of modes $n=9$,
$m=5$, $r=1$, $p=3$ with frequencies $-\Omega_9$, $\Omega_5$,
$\Omega_1$, $-\Omega_3$, respectively. One gets the $\Delta=0$ resonance
condition for the value $A=1.97$: the corresponding numerics and
analytical results are presented in Fig.~\ref{Fig2}(b).

\begin{figure}[t]
\epsfig{file=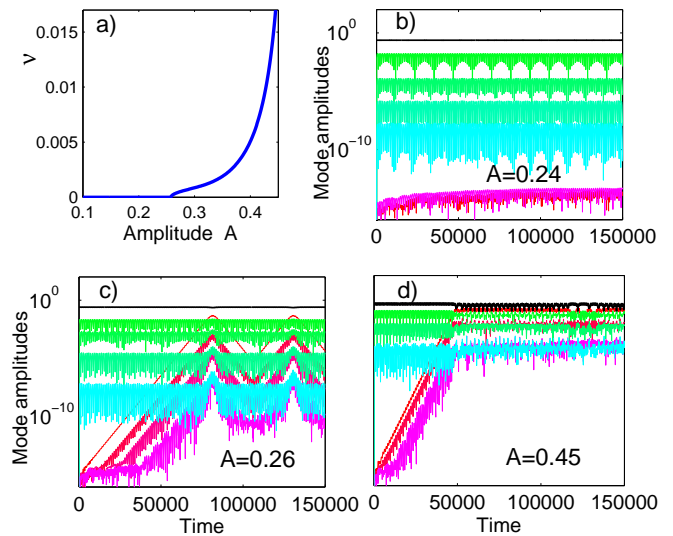,width=1\linewidth}
\caption{a) Analytically calculated growth rate according to
formula~\eqref{grow} in the case of the short-range FPU model. b) and c)
display mode stability/instability near the theoretically
predicted threshold value. d) shows mode dynamics far from the
instability threshold value. In all simulations, $N=32$.}
\label{Fig3}
\end{figure}

Now it is straightforward to extend the study of this unusual
instability behavior to the conventional short-range FPU model.
Indeed, mode evolution, Eq.~\eqref{first1}, is applicable for the
short-range case as well, the only difference is contained in the
unperturbed frequencies $\Omega_n^0$. Note that in the short-range
FPU unperturbed frequencies of low modes are almost equidistant, see
Fig.~\ref{Fig1}a., in contrast to the long-range case. Therefore,
the denominator in Eq.~\eqref{pert} is close to zero, which is the
reason for the well known phenomenon of FPU recurrences (large
energy transfer between the low modes)~\cite{FPU,galavotti,tod}.
Thus, although the instability mechanism is present in short-range
as well, the energy exchange among modes does not allow steady
exponential growth of the unexcited modes. The short-range FPU
instability develops completely only for larger values of the
amplitude $A$, when instabilities are not {\it sporadic}.

Seeking for the resonant modes in the short-range FPU model, one finds the set $n=2$, $m=N-2$, $r=1$,
$p=3$ with frequencies $\Omega_2$, $\Omega_2$, $\Omega_1$,
$3\Omega_1$, respectively. Then the condition $\Delta=0$ is reached
for $A=0.29$. Plotting again the growth rate~$\nu$ versus $A$ according to the
formula~\eqref{grow}, we get the dependence presented in Fig.~\ref{Fig3}a.
The instability is no more {\it sporadic}: instead, after it is
triggered at a given value of $A=0.249$, it persists
for all larger values. In Figs.~\ref{Fig3}b-d, we plot the modes
amplitude evolution for different values of $A$. Instability develops in the
range $A=0.24-0.26$, in excellent agreement with theoretical predictions.


Finally, as concluding remarks, we address the question of
estimating the threshold amplitude $A_{th}$ below which no sporadic
instability takes place in the large $N$ limit. The total number of
harmonics in the $N$ modes (see e.g. Eq.~\eqref{pert} for the third
mode) scales as $N! \sim N^N$; consequently the distance between
neighboring harmonics decreases as $N^{-N}$. Therefore, the average
detuning $\Delta$ in formula~\eqref{grow} should be of the same
order and thus $\Delta^2\sim N^{-2N}$. On the other hand, the first
term under the square root scales as $(A_{th}^2/N^4)^{N}$ in the
perturbative limit. From the condition that the growth rate in
Eq.~\eqref{grow} is real it follows that $A_{th} \sim N$ (this estimate
is compatible with recent unpublished numerical experiments~\cite{newexp}).
Therefore, the energy density threshold remains finite at large $N$,
$\epsilon_{th}=E_{th}/N \sim A_{th}^2/N^2 \sim const.$.

\begin{acknowledgments}
We acknowledge useful discussions with Andrea Trombettoni and we
thank George Chechin and Stepan Shcherbinin for providing us some
unpublished results on odd-even instability in the FPU model. This
work has been partially supported by the joint grant EDC25019 from
CNRS (France) and SRNSF (Georgia) and contract LORIS
(ANR-10-CEXC-010-01). TD and SR thank the Galileo Galilei Institute
for Theoretical Physics, Florence, Italy for the hospitality and the
INFN for partial support during the completion of this work.
\end{acknowledgments}


\begin{thebibliography}{99}
\bibitem{second} The second term in~\eqref{RJ} is chosen in view
of ensuring periodic boundary conditions and translational
invariance of the model, which enables us to diagonalize the harmonic
interaction in momentum space. This term vanishes in the thermodynamic limit
$N\rightarrow\infty$ and so has little relevance in analytical
considerations but could have a significant impact on numerical
simulations, where achieving this limit is impossible.
\end{thebibliography}
\end{document}